\documentclass[]{aastex631}

\shorttitle{Solar and stellar flares: frequency,
active regions and  stellar dynamo}
\shortauthors{Katsova et al.}
\graphicspath{{./}{figures/}}

\begin{document}

\title{Solar and stellar flares: frequency,
active regions and  stellar dynamo}

\author{M.M.Katsova}
\affiliation{Sternberg State Astronomical Institute, M.V.Lomonosov Moscow State University  \\
Universitetskij prosp. 13, 119991, Moscow, Russia}

\author{V.N.Obridko}
\affiliation{IZMIRAN,  4 Kaluzhskoe  Shosse, Troitsk, Moscow, 142190}
\affiliation{Central Astronomical Observatory of the Russian Academy of Sciences at Pulkovo, St.Petersburg, Russia}

\author{D.D. Sokoloff}
\affiliation{Moscow State University,  Moscow, 119991, Russia}
\affiliation{IZMIRAN,  4 Kaluzhskoe  Shosse, Troitsk, Moscow, 142190, Russia}
\affiliation{Moscow Center of Fundamental and Applied Mathematics, Moscow,
119991, Russia}

\author{I.M.Livshits}
\affiliation{Sternberg State Astronomical Institute, M.V.Lomonosov Moscow State University  \\
Universitetskij prosp. 13, 119991, Moscow, Russia}
\affiliation{Department of Geography and Environmental Development, Ben-Gurion University of the Negev \\
P.O.B. 653, 84105, Beer-Sheva, Israel}

\begin{abstract}

We demonstrate that for weak flares the dependence on spottedness can be rather weak. 
The fact is that such flares can occur both in small and large active regions. 
At the same time, powerful large flares of classes M and X occur much more often in large active regions. 
In energy estimates, the mean magnetic field in starspots can also be assumed equal to the mean field in the sunspot umbra. 
So the effective mean magnetic field is ~900 Mx/cm$^2$ in sunspots and ~2000 Mx/cm$^2$ in starspots. 
Moreover, the height of the energy storage cannot be strictly proportional to A$^{1/2}$. 
For stars, the fitting factor is an order of magnitude smaller. 
The analysis of the occurrence rate of powerful solar X-ray flares of class M and X and superflares on stars shows that, 
with allowance for the difference in the spottedness and compactness of active regions, both sets can be described by a single model. 
Thus, the problem of superflares on stars and their absence on the Sun is reduced to the problem of 
difference in the effectiveness of the dynamo mechanisms.

\end{abstract}



\section{Introduction} \label{sec:intro}

Solar flares are a spectacular phenomenon in the solar magnetic activity.
They can more or less directly affect the Earth, and the study of
solar flares is of both applied and academic interest. The
origin of solar flares is obviously associated with the solar magnetic
field and, in this sense, it is related to the action of the solar dynamo 
responsible for the formation of the magnetic field. The study of solar 
flares, which can be considered a traditional part of solar physics, 
has made impressive progress \citep[among many others, see, e.g.][]{P63, 
PF02, B08, BG10, K11, Eetal12, Setal12, S13, Aetal14}.

Of course, phenomena similar to the solar flares and known as
stellar flares occur on various stars. As is known, the total
energies of solar flares vary in a wide range of $10^{24}$-$10^{32}$ erg from the 
weakest events to the strongest ones \citep[see, e.g.,][]{Zetal20}.   
Stellar flares are
best studied on low-mass, red dwarf stars, and their total energies
exceed the maximum solar value by several orders of magnitude 
\citep[see, e.g.,][and references therein]{Hetal21}.
Besides, the most powerful of these flare phenomena were mainly recorded on very young dwarfs, including T Tau stars, members of open clusters, fast-rotating subgiants and giants, as well as on chromospherically active components of RS CVn-type close binaries \citep[see, for instance,][]{GAetal03, Fetal04,  Schmetal19}.

It is known that the most powerful flares on the Sun are rare phenomena that are characterized by sudden rise in optical continuum emission, and they are called as a white-light flare. Since a source of the flare optical continuum emission has a low contrast against the photosphere, a small flare area, and lives a short time (a few minutes), this prevents to reveal the temporal profile of the flare radiation. Nevertheless, powerful Sun-as-a-star flares were detected in long-term data on Total Solar Irradiance (TSI) \citep{K11}.   
Note that same problems make difficult the detection of optical flares on single main sequence G stars,  so it was thought that white-light flares have not been seen there until the Kepler mission. The only recently a definite information appears about flare activity of these stars \citep{Jaetal18,  Ketal21, Betal21}. From the other side, \cite{Kaetal21} showed that time profiles of solar flares in the UV-continuum emission are similar to impulsive flare light curves registered on the red dwarf during the Kepler mission. This result supports a point that an existence of optic flares in G stars can be suspected in the UV data. Indeed, they were found recently in GALEX NUV-data for these stars \citep{Betal19}.

It seems natural to use the ideas gained from the study of solar flares to understand stellar flares as similar
phenomena in the stellar physics. However, as shows the
progress in observations of the stellar activity, the stars
reasonably similar to the Sun can produce flares substantially more energetic than the strongest solar ones.

The total energies of the strongest stellar flares can
exceed $10^{36}$ erg in the optical wavelength range \cite[see, e.g.,][]{Hetal21}. As for the flare
activity of the solar-type  main sequence G stars, there was very little
information prior to the Kepler mission \cite[e.g.][]{Jaetal18, Koetal21, Betal21}. The superflare concept
applicable to powerful non-stationary stellar phenomena was
introduced when the first results of the Kepler mission, which operated in
2009-2018 and detected huge flares on G-type stars, were
published \citep{Metal12,Setal13, Metal15,Netal19, Oetal21}. These publications reported the detection of major stellar flares on solar-type stars with total energies
from $10^{33}$ to $10^{37}$ erg at optical
wavelengths. A more detail analysis showed that  most flares
had the total energy of $10^{33}$--$10^{34}$ erg \citep{Tetal21},  while only a
small fraction of the phenomena could be considered superflares
with  energies $E=10^{35}$--$10^{36}$ erg. Now, it is
clear that the most powerful events with $E> 10^{36}$ erg occur
either on components of the close binary stellar systems, or on subgiants and giants, or on very young  and/or fast rotating stars that have not reached the main
sequence \citep{B15, KN18, Tetal21}. 
The results of analysis of numerous multiwavelength observations of stellar flares and other non-steady phenomena on red dwarfs and solar-like stars reviewed by \cite{G05}  provides evidences  for their common physical nature with solar flares and confirm this idea first expressed  by \cite{GP72}. 
The process can be approximately described as a deposit of the free energy 
of the non-potential magnetic field in a certain volume, its impulsive 
release during non-steady event, and the subsequent response of the 
atmosphere to the resulting acceleration of particles and plasma heating. 
At the same time, already \cite{Getal87}  paid attention to the deficiency 
of up-to-date models of solar flares for explanation of the strongest 
stellar flares. In particular, it seems plausible that dynamo underlying 
magnetic in superflares stars is not fully identical to the conventional 
solar dynamo.

Indeed,
the maximum total flare energy on solar-type stars can be several times more than $3 - 5\times10^{34}\;$erg. 
This estimate is based on the magnetic virial theorem \citep{Letal15, KL15}. 
Similar value is discussed now in the recent statistics of all the primary Kepler mission data \citep{Oetal21}.
 
It looks plausible that magnetic configurations sufficient to accumulate corresponding magnetic energy have to be different in size and/or morphology from conventional sunspots.
 
Nevertheless, we believe that
the problem of superflares is, perhaps, less dramatic than it seemed earlier, however
still does exist and expects its explanation. 
We are going to propose corresponding revision in this paper.

To assess the similarity or difference between the solar and stellar flares correctly, it is necessary to take into account a number of circumstances.

First, a double selection of observational data must be taken into account. For natural reasons (the sensitivity of the equipment), only the most powerful white-light flares, which are extremely rare in the Sun (about 0.4\% of the total number of flares observed on the Sun over 15 years, see section 2 below), are recorded on stars. In addition, the rotational modulation technique makes it possible to detect only the largest concentrated spots or spot groups on stars. We discuss these issues in Sections 1 and 2.

Another circumstance that must be taken into account is that flares of different energy depend in different ways on the area of  the active region. The frequency and energy of weak X-ray flares B and C are virtually independent on the area of the active region and, therefore, cannot be used to assess the similarity or dissimilarity with superflares on stars. This issue is discussed in Section 2.

And finally, when evaluating the total magnetic energy in the active region, we cannot use the extreme values of several kG, which are observed in sunspots. These values correspond to a very small part of the spot. To find the total energy, it is necessary to obtain the integral values, for which we have to know the distribution of the magnetic field over the active region. In this case, we cannot use the photometric values of the spot area, but have to introduce the concept of a magnetic boundary and, additionally, to determine the relative fraction of the umbral area. These estimates are given in Section 3.

These problems were discussed by various authors: see, for example \citep{B08, BG10}. A particularly detailed and thorough analysis was carried out by \cite{B05}, who not only described methods for studying starspots, but also provided extensive observational material, which was subsequently used by other authors \citep{Aetal13,  Setal13, Netal13, Netal19, Metal15, Hetal21, Oetal21}. It should be noted that the procedures used to determine the spot areas on the Sun and stars are essentially different. In the former case, the observer directly calculates the area of each spot from the full image of the Sun and, then, sums up the values obtained. The penumbra is traditionally included in the spot area. The procedure of determining the total spottedness on stars is more complicated. First of all, one has to find out the variation in the star brightness. The methods for determining this variation, such as the light--curve modeling, Doppler imaging, Zeeman--Doppler imaging, molecular bands modeling, are described in detail by \cite{B05}. These methods are based on the use of different radiation characteristics of a star, continuous spectrum in different ranges, different spectral lines, Doppler effect, magnetic splitting, and molecular spectrum. Generally speaking, these data can refer to different layers in the stellar atmosphere. Standing apart is the spot temperature. The large sunspot areas and temperature contrasts found in active stars suggest that the photometric and spectroscopic variability of these stars is dominated by the starspot umbra. Our current knowledge about starspot temperatures is based on measurements obtained from simultaneous modeling of brightness and color variations, Doppler imaging, modeling of molecular bands and atomic line-depth ratios, the latter being the most accurate method. A representative sample of starspot temperatures for active dwarfs, giants and subgiants is provided in Table 5 and plotted in Fig.~7 by \cite{B05}.
 
Then, the spot area ($A_{\rm spot}$) of superflare stars is estimated from the normalized amplitude of light variations ($\Delta F/F$) by using the following equation 

\begin{equation}
A_{\rm spot}=\frac{\Delta F}{F}A_{\rm star}\left[1-\left(\frac{T_{\rm spot}}{T_{\rm star}}\right)^4\right]^{-1}, \label{spot}
\end{equation}
where $A_{\rm star}$ is the apparent area of the star and $T_{\rm spot}$ and $T_{\rm star}$ are the temperature values of the starspot and stellar photospher.

No matter how the temperature values are obtained, it turns out that in solar-like stars they are close to 
the temperature in the sunspot umbra \citep[see][Table~5]{B05}. This means that, in fact, 
we find the total area of the umbra or, to be more precise, the starspot area on stars can be 
considered coinciding with the area of the umbra. Therefore,  in energy estimates, 
the mean magnetic field in starspots can also be assumed equal to the mean field in 
the sunspot umbra. 

Considering all of the above--mentioned circumstances, we arrived at a conclusion that the problem of superflares on stars and their absence on the Sun is reduced to the difference in the effectiveness of the dynamo mechanism. This conclusion and certain consequences for the problem of the generation of magnetic fields in the Sun and stars are discussed in Section 4.

\section{Comparing solar and stellar flares} \label{Comp}

The structure of our research depends substantially  on the
particular difficulties of comparing the solar and stellar flares
listed in this section.

First of all, we have to emphasize that the task of comparing the stellar and solar flares is far from straightforward. Indeed, it is often stated that \cite[see, e.g.][]{Hetal21}
solar flares do not fit into the linear trend visible in stellar data. The
point, however, is that due to the instrumental limitations, the
stellar flares are only white-light flares, which are the most
energetic ones, while the weak flares are not observable and, therefore, are absent on the plot. In contrast, the solar data
contain all flares with peak X-ray flux from $10^{-7}$ W/m$^2$ up to the flux of the order of $10^{-3}$ W/m$^2$, i.e., from subflares up to
the major proton events. The most energetic flares belong to 
classes  M and X. This makes up the energy range from $10^{28}$ to $10^{32}$ erg and corresponds to a spottedness of no more than 3000 m.v.h.  The minimum detectable spottedness on stars is approximately 1000 m.v.h. \citep{Oetal21, Oetal21a}, but in general stellar flares have energies from $10^{34}$ to $10^{36}$ erg, which corresponds to the spottedness range from 0.01 to 0.3 of the area of visual solar hemisphere. 

Another relevant problem connected with the fact that the total spot area (spottedness) is not the only parameter that determines directly the flare energy. This point is discussed in Section 3.

\begin{figure}
\vspace*{0.2cm}
\begin{center}
    \includegraphics[width= 0.9 \columnwidth]{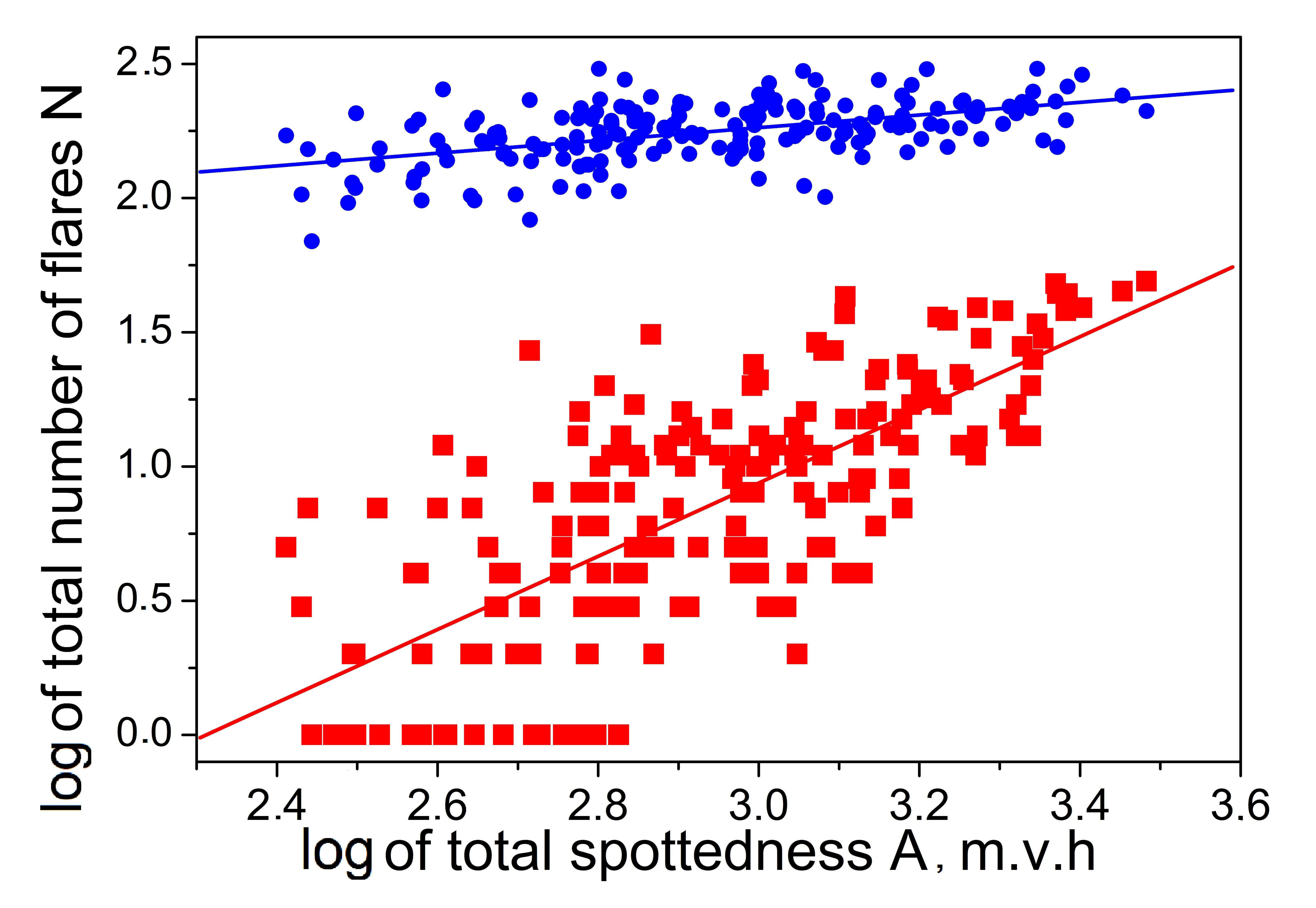}
       \caption{Monthly mean occurrence rates of flares N versus monthly mean spottedness A (top, blue circles -- B and C classes flares;  bottom, red squares -- M and X class flares data). The colored lines show the result of a linear
regression of each of the two samples of flares}
    \label{fig1}
\end{center}	
\end{figure}

Note also that the expression "spot area" or "spottedness" is
understood quite differently when referred to the Sun and stars. As to
the Sun, we assume that the spottedness is just the total area of
individual spots visible in the white light relative to the
area of the solar hemisphere. In contrast, the stellar spottedness is
usually estimated from the rotation modulation of the stellar brightness without taking into account the spot distribution over the star surface. This method, however, gives basically different estimates for a single large spot and for many spots of moderate size distributed more or less homogeneously over the stellar surface.   In particular, when observing the Sun as a star by this method we see almost no rotational modulation even in the periods of very high activity. 

Summarizing all said above, we expect a strong selective effect
in stellar observations, which gives preference to the contribution of
a single or a few very large spots. Therefore, the solar
data have to be properly selected to be comparable with the stellar
ones.

It can be expected that for weak flares the dependence on spottedness can be rather weak. The fact is that such flares can occur both in small and large active regions.
At the same time, powerful large flares of classes M and X (the peak flux larger than $10^{-5}$ W/m$^2$) occur much more often in large active regions. It is known that there are positive correlations between
the sunspot coverage and the energy of the largest solar flares \cite[see, e.g.][]{Setal00}.

To test these considerations, we performed an additional analysis. The occurrence rates N of flares of different classes were estimated for the period 1992--2016 using data from  the catalogue $http://hec.helio-vo.eu/hec/hec\_gui.php$, which contains 44566 X-ray flares. GOES Soft X-ray Flare List was used. The flare classes are determined by their peak fluxes as follows:
B stands for fluxes $(1-9) \times 10^{-7}$ W/m$^2$ (17747
flares), C stands for $(1-9) \times 10^{-6}$ W/m$^2$ (24190
flares), M corresponds to  $(1-9) \times 10^{-5}$ W/m$^2$
(2449 flares), and X stands for  $(1-9) \times 10^{-4}$ W/m$^2$
(180 flares). Based on these data, we calculated the monthly mean occurrence rates of flares of each class and compared them with the monthly spottedness data taken from
$https://solarscience.msfc.nasa.gov/greenwch/sunspot\_area.txt$. Then,
we gathered separately the data for classes  B and C, and classes M and X and plotted them versus the spottedness (Fig.~1). For weak flares (B and C), the occurrence rate $N_{BC}$ is almost independent of spottedness (Fig.~1). The exponent is only 0.237, correlation coefficient 0.514 . 

\begin{equation}
\log N_{BC}=1.55185 \pm 0.0845 + (0.23674 \pm  0.0286)\log A \, .
\label{eq2}
\end{equation}

A pronounced relationship between the occurrence $N_{MX}$ and the spottedness $A$ is seen only for strong flares of class M and X (Fig.~1, bottom). The exponent of the power-law dependence for these
flares is much higher and amounts to 1.363, correlation coefficient 0.728. 

\begin{equation}
\log N_{MX}=-3.152 \pm 0.2749 +(1.36349 \pm 0.09318)\log A \, .
\label{eq3}
\end{equation}

\begin{figure}
\vspace*{0.2cm}
\begin{center}
    \includegraphics[width= 0.9 \columnwidth]{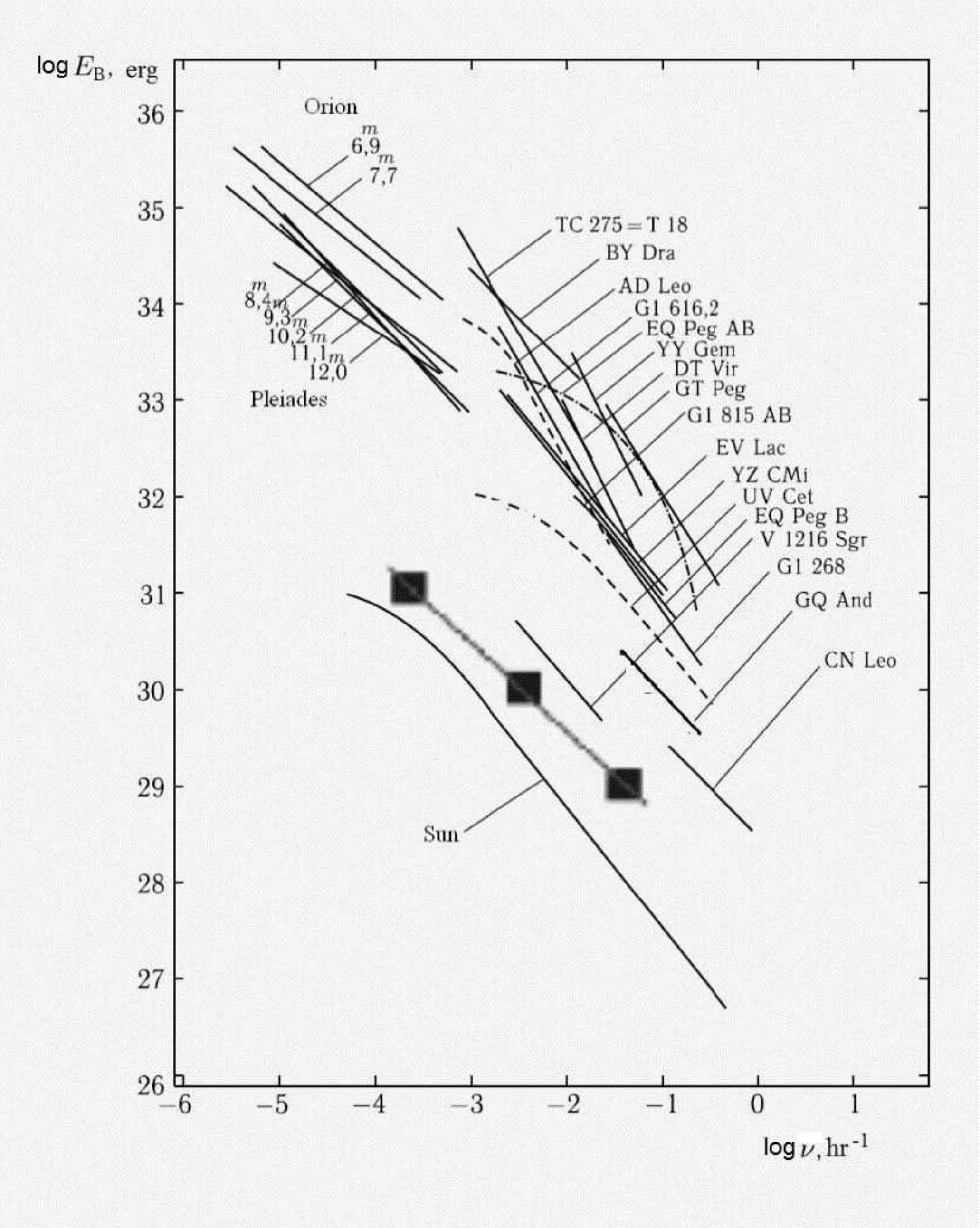}
\vspace*{0.2cm}
       \caption{The  energy of flares in connection with their occurrence rates on different stars. The curve for the Sun is based on $H_{\alpha}$ data, the  squares and thick line are plotted according to X-ray data.}
    \label{fig2}
\end{center}	
\end{figure}

An important characteristic of flares is their occurrence rate. The simplest and physically meaningful
relationship between the flare energy and occurrence rate is expressed by a power function. Fig. 2 based
on the picture from \cite{Getal87} repeated later by \cite{G05}
shows the
relationship between the energy of flares observed in the photometric $B$-band ($E_B$) and their occurrence rates for
different dwarf stars, indicated on the plot, and for the stars of Pleiades and Orion, as well.  These data are the result of thousands of hours of patrol photoelectric observations at various world observatories. By now, more than 3000 stars of the type under discussion have been registered. The data shown in the figure refer mainly to red dwarfs (the spectral class M). The total energy ranges from 10$^{28}$ to 10$^{36}$ erg, which is much broader than the superflare range (10$^{33}$--10$^{36}$ erg). The energy range of superflares was determined from observations of the Kepler mission, which could not discriminate weak flare because of the background noise. More recent data obtained with the Transiting Exoplanet Survey Satellite (TESS) and Large Sky Area Multi-Object Fiber Spectroscopic Telescope (LAMOST) are described by \cite{Tetal21}.The authors developed their own technique, which allowed recording weaker events than those detected by Kepler. \cite{Coletal22}, validated the procedure by comparing the results obtained with other techniques by \cite{Maretal21} for AU Mic and found a rate of events of $\approx5$ flares per day with energy $E_f > 5\times10^{31}$ erg. In addition to that, they studied the system DS Tuc A and found a rate of energetic events of $\approx2$ flares per day with energy greater than $2\times10^{32}$ erg. It is interesting to note that these data agree well with ground-based observations of red dwarfs (see Fig. 2).

It should
be noted that a good agreement between the observed relation and the power function, which is represented
by a straight line in the logarithmic diagram, does not exist in the entire energy range. Sometimes there
is saturation for very strong flares and a sudden dip due to the observational selection for very weak
flares \citep{G05}. Therefore, in general, the figure shows only the linear sections of the relation.
Significant deviations from linearity are observed for UV Ceti, AD Leo and EQ Peg AB and are shown by
dotted lines. Additionally, the figure shows the relationship for solar flares in H$\alpha$. The total length of the dynamic energy range of flares on each star does not exceed two orders of magnitude.

The range of the energies and occurrence rates on the diagram is rather broad (7--9 orders of magnitude), 
while the angular coefficients of the dependencies $\beta=d\log\nu /d\log E$ lie in a narrow range from 
--0.5 to --0.9. In the clusters, they are somewhat larger in the absolute value (from --0.8 to --1.0); 
for solar flares in H$_{\alpha}$, $\beta=-0.8$ \citep{G05}. We have plotted a similar curve using 
observations of X-ray flares mentioned above. The occurrence rates are 0.110 per hour for class C flares 
(24190 events), 0.0112 events  per hour for class M flares (2449 events), and $8.208 \times 10^{-4}$ 
per hour for class X flares (180 events). The dip for class B flares was ignored as it was when plotting 
the relations for starflares. The result is represented in Fig. 2 by a thick line and large squares. 
One can see that, here, the value of $\beta$ (-1.06) is close to the values for superflares in the 
clusters and is typical of stars of approximately the solar age \citep{G05}. The similarity of the 
values of $\beta$ shows that despite the significant difference in energy, superflares on stars 
and solar flares are apparently determined by the same processes \citep{G05}.

It should be noted that on the Sun, the linear part of the dynamic energy range also does not apparently exceed 2--3 orders of magnitude. If there were no saturation and with the observed value of $\beta$, at least one superflare with an energy in the range of $10^{34}$--$10^{35}$ erg would have to occur on the Sun every 550 years, which is not confirmed by the historical and archaeological data available.

Note also that, as follows from Fig. 2, flares of the same energy on the Sun (and apparently in general on G dwarfs) are 30--100 times less frequent than on red dwarfs.

To avoid misunderstandings, we note that the value of ${\nu}$ used by us, calculated per one hour, corresponds to the $f_\ast$ used by \cite{Tetal21}, which is calculated per a year, but does not coincide with the index $f_n$ introduced there, in which additional normalization is carried out for the number of observed stars and the energy range. This more physically sound parameter yields $\beta = -1.76$.

\section{Scaling stellar data}

Now, our task is to suggest a reasonable scaling for the energy $E$ of stellar flares. We depart from the conventional assumption that the energy is determined by the magnetic energy density $B^2/8 \pi$ (where $B$ is the magnetic field) and the volume of energy accumulation. The latter is proportional to the spot area $A$ and the height $h$ of the volume. 

We depart from the viewpoint that the initial flare energy release 
originates in the magnetic energy in a volume. This energy is converted 
in particle acceleration, optical and X-ray radiation, plasma motions  
\cite{Eetal04, Betal05, V12}. Immediate origin for the energy release 
in a flare is a current dissipation which s proportional with the scaling 
factor $f_r$ to the total magnetic field energy. Estimates of $f_r$ contain 
various uncertainties \cite{Setal12} and varies from 0.01 to 0.5. 
We argue that solar and stellar flares are physically similar and fortunately 
we can accept that $f_r$ is the same for solar and stellar flares. 
Giving below corresponding references we do not focus below attention on 
this scaling factor. 

Thus, the total energy released in a flare is described by the expression

\begin{equation}
E=f_r\int{\frac{B^2}{8\pi}}dV
\end{equation}

Direct calculations by this formula are difficult even for the Sun and 
require observations with good spatial resolution and high-quality 
full-vector maps of the magnetic field. Such calculations with some 
additional assumptions have been performed by several authors 
\citep[e.g., see][]{Letal15, Zetal20} and generally confirmed the 
above concept with the parameter $f_r\approx0.1$. However, 
such calculations are completely impossible today for stars and, 
therefore, a simpler formula is used:

\begin{equation}
E=f_r\frac{\overline{B^2}}{8\pi}V
\end{equation}
Here, $B$ is the mean field in the given volume, which is determined as follows:

\begin{equation}
V=\overline{A}\cdot\overline{H}
\end{equation}

Situation with another scaling factors $f_h$ and $f_s$ is more delicate 
as they may be substantially different for the Sun and stars. The point 
is that estimating $B$ we can use magnetic field averaged over the whole 
sunspot while for the stars we have to use magnetic field averaged over 
the umbra as  magnetic field in the whole star spot is not accessible 
for observations. The total sunspot area $A$ for the Sun is determined 
photometrically and includes sunspot umbra and penumbra. What about stars, 
the value $A$ is determined using Eq.~(1) basing on spectral temperature 
related to the umbra \cite{B05, Hetal21}. We need the scaling factor $f_s$ 
to reproduce this difference.

\begin{equation}
\overline{A}=f_s A_{\rm spot}
\end{equation}

The value of $\overline{B}$ changes accordingly. For the Sun, we have to use the mean field value over the entire sunspot, and for the stars, the mean field in the spot umbra.

And finally, to estimate $\overline{H}$, we use the expression connecting the height of the energy release region with the radius of a round spot:  

\begin{equation}
H=f_h A^{1/2}_{\rm spot}
\end{equation}

Note that this approximation raises serious doubt. Here, we also need to introduce a model parameter $f_h$, since the region of primary energy release must contain free (nonpotential) energy. Deviations from potentiality arise when the pressure and energy of plasma motions are greater than or comparable to the potential energy of the magnetic field. In the majority of solar flare models, the height $H$ is 10-20 thousand kilometers, i.e., is comparable to the radius of a very large sunspot, but is much smaller than the radius of a stellar spot. 

Without this additional parameter, very large values of $A$ will yield too large $H$; e.g., at $A=0.1$, we will get a height comparable to the radius of the star, which, obviously, cannot give any reasonable current density. This means that, at equal heights of the energy release region, the parameter $f_h$ on the stars is smaller.

The other drawback of this approximation is that $\overline{H}$ is not an additive parameter. If several large spots are observed on the star, their area is summed up, and the parameter $\overline{A}$ increases proportionally, while the value of $\overline{H}$ does not change.

Combining these formulas, we obtain our Eq. 9

\begin{equation}
E = f_h\cdot f_s\cdot f_r (B^2/8 \pi) A^{3/2}. 
\label{eq4}
\end{equation}

When comparing with observations, it is usually not taken into account that the magnetic field strength and the dimensionless fitting factors may change as the spot coverage changes \citep[e.g., see][]{Netal13, Metal12, Metal15, Oetal21}.  Note that the above equation is nearly the same as Eq.~(9) by \cite{Netal13} however we introduce scale factors while \cite{Netal13} use observed quantities combined with a reasonable estimate of the ratio of the spot temperature to the stellar photospheric temperature.
We use the scaling factors to take into account that solar and stellar data are obtained using substantially different methods. 

Therefore, on the diagrams in logarithmic coordinates this dependence is represented as a straight line, and comparison with observations is carried out by selecting a constant value of the magnetic field. The usual conclusion is that it is not possible to find a constant field value at which Eq. 9 would describe both solar and stellar flares. This does not take into account that both the B value and the fitting factors can be different on the Sun and stars.

As mentioned above in the Introduction, in sunspot observations, the photometric picture is used to calculate the entire area of a sunspot including the penumbra. The sum of these values is defined as the spottedness and is contained in all solar catalogs. For stars, a different procedure is used, which is based on the temperature difference between the star and the observed spot. This difference corresponds to the temperature difference between the spot umbra and the star. This means that, in fact, we find the total area of the umbra or, to be more precise, the area of a starspot can be considered coinciding with the area of the umbra. Therefore, in energy estimates, the mean magnetic field in starspots can also be assumed equal to the mean field in the sunspot umbra.

Thus, to estimate the parameters included in Eq.~(9), it is
necessary to know the mean magnetic field in a spot. However, the sunspot boundaries are determined based on photometric properties. Unfortunately, there is still no generally
accepted definition of the magnetic boundary of a spot. In this
work, we used SDO/HMI observations to solve this problem.

We considered the daily SDO/HMI data on the line-of-sight magnetic field component for the period from May 1, 2010 to October 31, 2016 -- the total of 2375 days. The daily data on sunspot numbers were downloaded from the WDC-SILSO website, Royal Observatory of Belgium, Brussels http:
//sidc.oma .be / silso / datafiles (version 2). The total daily sunspot
areas were taken from the NASA website https://solarscience.msfc.nasa.gov/greenwch.shtml. 
The daily values of the line-of-sight field component were
recalculated into the radial component by dividing by the cosine of
the position angle. The area of each pixel was also corrected.
Then, we calculated the relative fraction $S_B$ of the area occupied by fields above a
certain limit. This fraction was expressed in
millionths of the solar hemisphere, as is customary when studying
the total areas of sunspots.

These daily values are expressed in m.v.h. of the hemispheric area and are
calculated for several thresholds ranging from 0 to 1800 G. As a first
approximation to finding the  magnetic field boundary, we calculated the regression between $S_B$ and the total sunspot area. It turned out that, at the magnetic spot boundary of 550 G, the
correlation between these values reaches 0.98. Moreover,
this correlation is valid in a very narrow range; even at the threshold values  of 500
and 575 G, the correspondence deteriorates. 

The calculation procedure is described in more detail by
\cite{OS18
}. Close values for the magnitude of the vertical component of the magnetic field at the outer boundary of the penumbra are also given by \cite{KM96, Setal06, Aetal13, BI11}

Assuming that the boundary of the spot area responsible for a flare
corresponds to the magnetic field of 550 G and plotting 
 the mean magnetic field $B_s$ in the spot versus
the spottedness (Fig.~3a), we learn that the spottedness $A=300$
m.v.h. (i.e. $3 \times 10^{-4}$ of the area of the visible solar hemisphere)
corresponds to $B=800$ G, while $A=900$ m.v.h. gives $B=900$ G. Taking
into account that the area of the solar hemisphere is $3.044 \times
10^{22}$ cm$^2$, we obtain from Eq. 4 the following lower
estimates for the total magnetic energies stored in sunspot regions with $A=300$ m.v.h and $A=900$ m.v.h.: $E = 8 \times
10^{32}f_h f_s f_r$ erg  and $E = 5.8 \times 10^{33}f_h
f_s f_r$ erg correspondingly.

\begin{figure}
\vspace*{0.2cm}
\begin{center}
  \includegraphics[width=0.85\columnwidth]{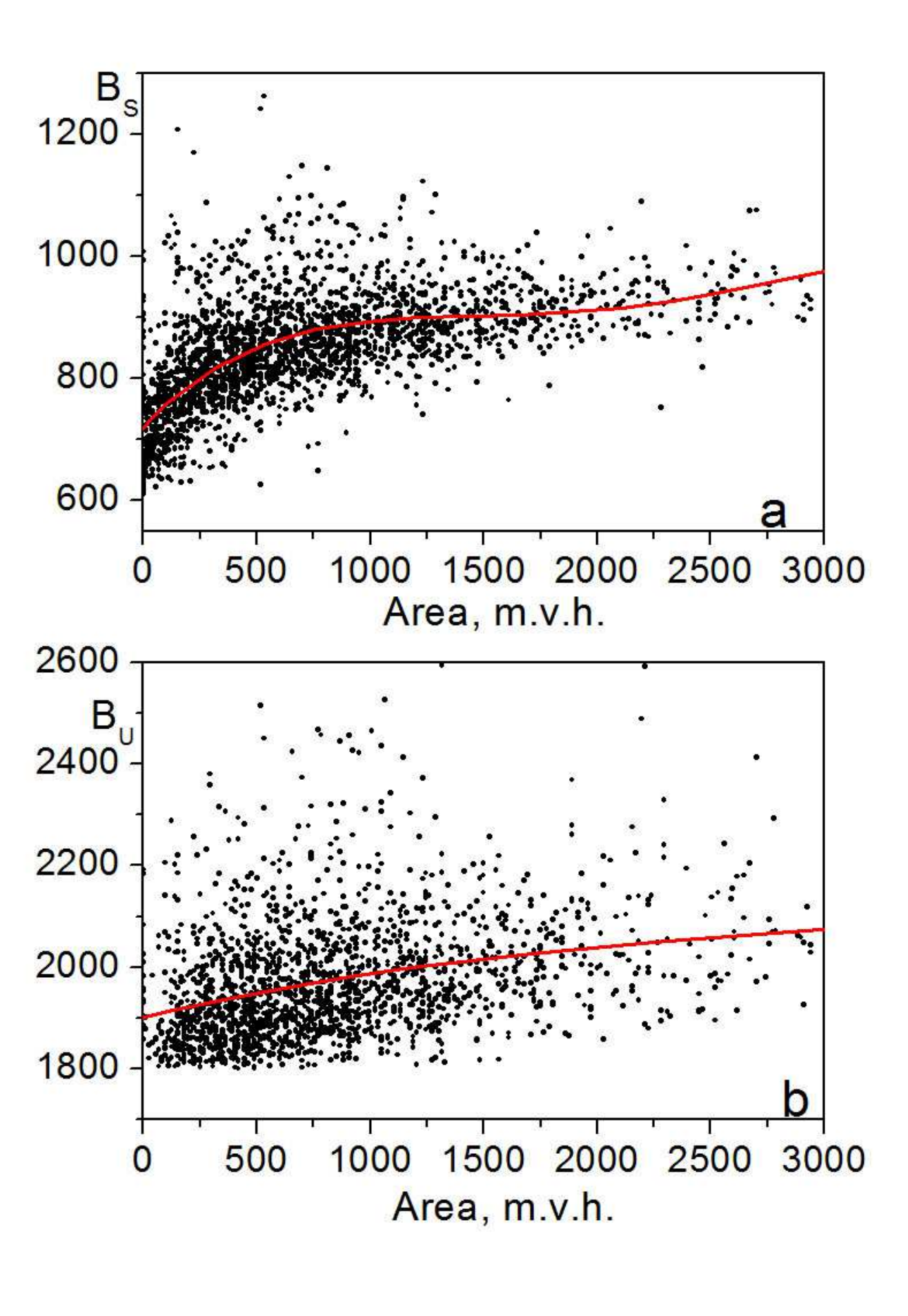}
  \caption{Mean magnetic field  in the flaring sunspots (a) and starspots (b) versus spottedness. 
           The magnetic field is 550 G at the sunspot boundary and 1800 G in the starspot boundary.}
    \label{fig3}
\end{center}	
\end{figure}

When making similar calculations for stellar flares, we have to take into
account that the spottedness $A=3 \times 10^{-2}$ m.v.h. is rather high, and the starspot must be more compact than the sunspot to be distinguished by the rotation modulation of the stellar
brightness. 

As mentioned above, the method of determining the area from temperature data leads to the fact that the determined area of the spot is essentially the area of the umbra. Therefore, under the assumption that the average field in the umbra is the same in sunspots and starspots, it can be calculated based on the knowledge of the magnetic field at the umbra--penumbra boundary. Unfortunately, the method that was used above to determine the outer boundary of the spot could not be applied due to the lack of a database on the sum of umbra areas on hemisphere. There are a number of works in which this umbra--penumbra boundary is discussed on the basis of direct observations \citep {J11, Jetal15, Jetal17, Jetal18, Setal18,  Letal20}.

In this work, we have chosen the value of 1800 Mx/cm$^2$ as the umbra boundary, which is quite close to the results of \cite{J11, Jetal15, Jetal17, Jetal18}. So we obtain the average magnetic field of the umbra 2000 Mx/cm$^2$ (see Fig.~3b).

As mentioned above, the temperature in stellar spots corresponds to the temperature of the sunspot umbra. Therefore, Fig.~3b shows a diagram for starspot umbra.

After having estimated the mean magnetic field $B$ in sunspots and
in starspots, let us estimate the scaling factors $f_h$, $f_s$, and
$f_r$.

Assuming that the mechanism of solar and stellar flares is the same, this height should be approximately conserved. To determine the dependence of this height on spottedness, the factor $A^{1/2}$ is provided in formula (9). It is easy to see that this multiplier gives too large values and reflects only the general trend. So for areas $3 \times 10^{-4}$, $10^{-3}$, $10^{-2}$ and $10^{-1}$ of the solar hemisphere, we get the values of $A^{1/2}$ equal to $30$, $50$, $170$ and $550$ Mm. In the Sun, the estimated height of the energy release domain is $10-20$ Mm \citep[see e.g.][]{Shetal18, Setal20, Zetal20}. Therefore, for the Sun, we must take the parameter $f_h = 0.3$, while in superflaring stars  we obtain $f_h = 0.1$.

$f_s$ is a dimensionless scaling factor determining the share of the
region occupied by the strongest magnetic field. This is actually
the relative area of the umbra in a sunspot. 
We assume $f_s$=0.2 \citep[see][]{Betal14}.

When estimating $f_s$ for the stars that produce superflares, we have to take into
account that the spottedness $A$ is expected  to be large. In
principle, a large spottedness can be achieved by increasing either the
number or the size  of spots. In the former case, however, 
the effect will be undetectable by observations based on the rotational modulation. Therefore, we have to assume that the observed stellar spottedness is determined by the relative area of the umbra of a single
large stellar spot. In other words, for stars with superflares we have to assume $f_s = 1.0$.

$f_r$ is part of the magnetic energy converted to radiation during a
flare. When estimating $f_r$, one has to be more careful. \cite{Hetal21},  following  \cite{Setal12}  who, in turn, followed \cite{Metal05, Setal08}, obtained $f_r = 0.01-0.5$ and based the width of the fitting strip on this estimate. Below, we will comment on this estimate in more detail.

Variations in the photospheric magnetic field in strong solar flares was investigated by  \cite{SH05, PS10, Maetal12}. Recently, \cite{CDetal18}  analysed 77 solar flares and found out that most major flares (class above M1.6) were accompanied by abrupt and permanent variations in the photospheric magnetic field.  They
considered 38 X class flares and 39 M class flares. For each flare, they isolated an area in the corresponding active region where the field variation lasted as long as 15 min. The amplitude of the field variation ranged
from the lower observational limit  of about $10$ G  to about
$450$ G (two cases). The mean amplitude was $69$ G.
In the case of X class flares, the variations used to be substantially larger than in the case of M class flares. The authors insist that the above amplitude estimates are representative.

\cite{Letal15, Shetal18, Setal20, Zetal20, Aetal21} estimated the ratio of the free and total energy as $0.15-0.25$. The results depend on the extrapolation method. However, only part of the free energy (from percents to dozens of percents) can be spent to create a flare. The volume of the flare occupies part of the active region only. During a flare, the energy can even grow somewhere inside the active region. The components of the photospheric magnetic field  can grow stepwise  during the flare \cite[e.g., see][]{P13, Setal17} for the horizontal component and \cite{PS10} for the line-of-sight component). On the one hand, the buoyancy can transport the magnetic flux in the flaring region even during a flare, while on the other hand, the flux can trigger the flare. There are flare models that take into account the energy income during the flare \citep[e.g.,][]{Moetal05, MS06}.

All these factors result in a 2-3-order difference in the 
flaring power and, thus, account for C, M, and X flares. In principle,
one can estimate $f_r$ for individual flares or flares of particular
types. Here, however, we are interested in the general link between
the energy and spottedness. Therefore, we did not use the value of 
$f_r$ in further calculations, since $f_r$ and, to some extent, 
$f_s$ cannot be reliably determined from observations and they are 
not a general characteristic of the flare phenomenon. They determine 
the energy of each individual flare, leading to a huge scatter 
in the observed values.

\begin{table}
\large
\caption{Model parameters}
\begin{tabular}{|c|c|c|c|c|c|c|}
\hline
$A$, m.v.h. & $B$, Mx$\cdot$cm$^{-2}$ & $f_h$ & $f_s$ & $f_r$ & $\log E$ & $H$, km \\
\hline
-3.5 & 800 & 0.3 & 0.2 & 0.1 & 30.659 & 9302 \\
-3.0 & 900	 & 0.3 & 0.2 & 0.1 & 31.511 & 16543 \\
-2.0 & 2000 & 0.1 & 1.0 & 0.1 & 33.926 & 17438 \\
-1.5 & 2000 & 0.1 & 1.0 & 0.1 & 34.676 & 31009 \\
-1.0 & 2000 & 0.1 & 1.0 & 0.1 & 35.426 & 55144 \\
\hline
\end{tabular}
\end{table}

A summary of the model parameters we adopted and the calculated values of energy E and flare height H are shown in Table 1 and Figure 4.  A cloud of points from \cite{Metal15} is partially copied to the same figure for energies above $10^{30}$ erg and spottedness above $10^{-4}$ of the solar hemisphere.  It can be seen that the values obtained by us generally agree with the observations. This figure is consistent with Figs.~4 and 5 in \cite{Oetal21} for Sun-like stars.

We conclude that the solar and stellar flares can be considered within the framework of a unique approach with specific governing parameters applied in the particular cases.

\section{Discussion and Conclusions}

Naturally, we (as well as all other authors we cite) assume that spots on the Sun and stars are of the same nature. In this case, the transformation of the rotational modulation into the area of the spot (or, more precisely, the umbra of the spot) using Eq. (1) is quite natural and provides a basis for comparing the dependence's of the number of flares on the spottedness. If we assume that sunspots and stars have different structures, then the very comparison of flare activity with spottedness loses its meaning and needs to be fundamentally revised.

 We have demonstrated that the mechanisms of solar flares and stellar superflares are basically identical and the corresponding data can be described by Eq.~(9) with realistic parameters. A compact solar active region with the umbral area of the order of 0.1 of the solar hemisphere and the magnetic field $B = 2$ kG (which gives the magnetic field strength of about $100$ G after averaging over the whole stellar surface) can produce a superflare with $E = (1-3) \times 10^{36}$ erg.

Thus, the flare generation mechanism can be the same on the Sun and stars.  
The main difference is that the spottedness on the Sun is no more than a few thousand m.h.v., while on the stars it reaches tenths of the area of the disk. As a result, the variation in the solar optical radiation is less than 0.1\% \citep{Fr06, Fr12} and on M dwarfs it reaches 10\%. At the same time, flares on the Sun are 1-2 orders of magnitude less frequent than on these stars and their energies do not exceed $3\times10^{32}$ erg. The mean magnetic field in sunspots is lower by a factor of 2 than in giant spots associated with superflares. Accordingly, the magnetic flux on M dwarfs is 3 orders of magnitude higher than on the Sun.

The problem is why the solar dynamo produces magnetic fields
associated with active regions of about $10^{-3}$ of the area of the
solar hemisphere, i.e. by $2-3$ orders of magnitude smaller than
required to get a superflare (about $A=0.1$ of the area of
the stellar hemisphere). Note that earlier \cite{Ketal21}   admitted the possibility that superflares are governed by a physical mechanism basically different from the solar mechanisms.
A comprehensive quantitative model which accumulates physical processes from dynamo action in stellar interior up to the flare formation is far above contemporary theoretical abilities even for the Sun not to say about another stars. Here we discuss some hints concerning this future theory which can be associated with superflares. 
Perhaps, to explain very high spottedness of superflaring stars one may to assume that the dynamo action domain on such stars is located just beneath the surface of the star rather than somewhere near the bottom of the convection zone, e.g., in the overshoot layer as it is generally believed.

\begin{figure}[t]
\vspace*{0.2cm}
\begin{center}
    \includegraphics[width=0.9 \columnwidth] {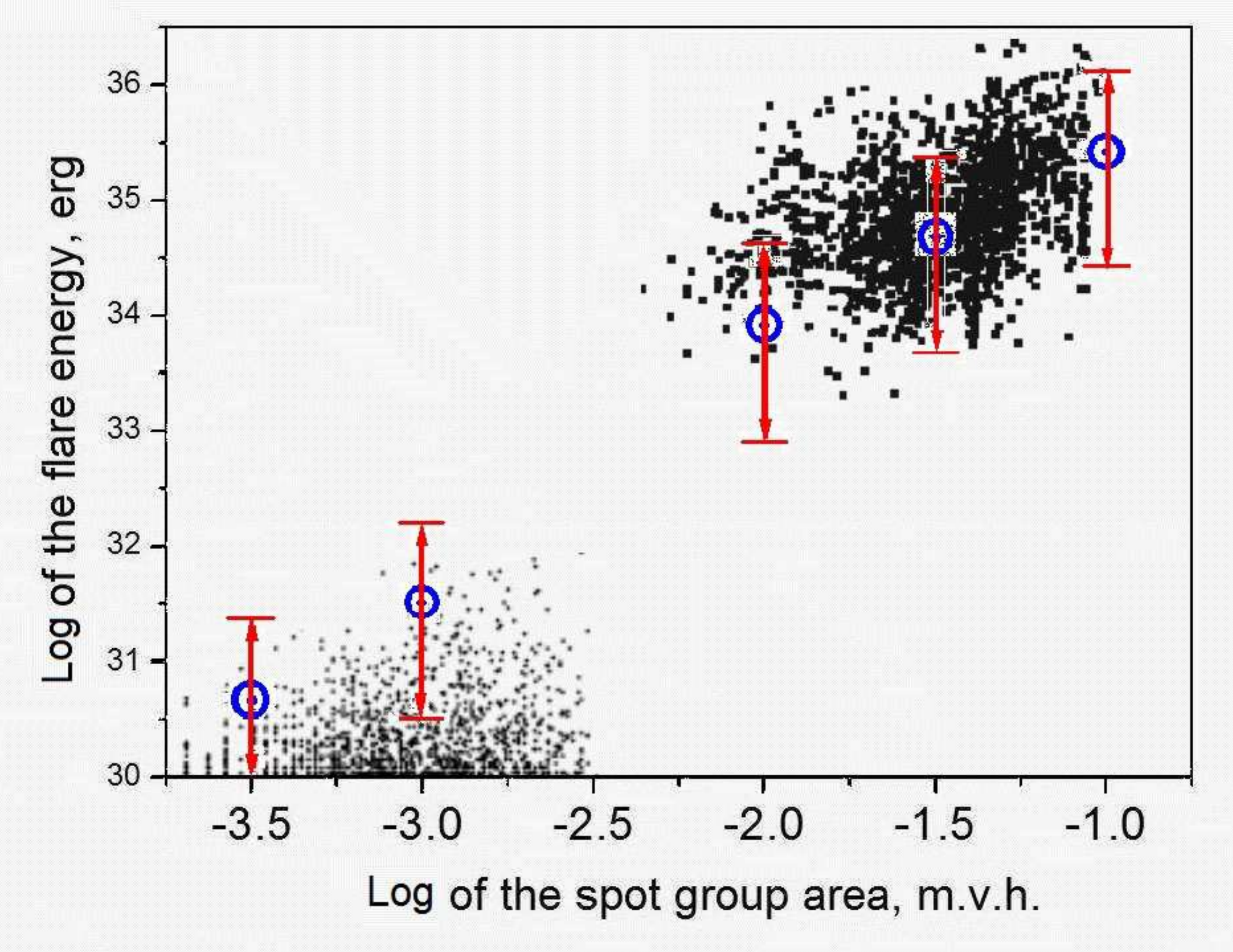}
       \caption{Flare energy versus total spot group area. A cloud of points 
	   from \cite{Metal15} is partially copied to the same figure for energies 
	   above $10^{30}$ erg and spottedness above $10^{-4}$ of the solar 
	   hemisphere. Results of our calculations are shown by blue circles. 
	   The dispersion of energy values associated with the uncertainty of 
	   the parameter $f_r$ (0.5 -- 0.01) is shown in the figure with red arrows.}
    \label{fig4}
\end{center}
\end{figure}

Indeed, helioseismology data indicate that the solar convection zone
contains two layers of substantially differential rotation. One is the so-called overshoot layer near the bottom of the convection
zone, while another is located near the solar surface. There are
solar dynamo models with dynamo action concentrated in the upper
layer of differential rotation  \citep[say,][]{Br05}; however, the
models with deep location of the dynamo active region 
are more
popular.

In order to produce an area with a strong magnetic field on the solar
surface in the form of a sunspot, the dynamo driven magnetic field has to propagate through a thick dynamo inactive layer. The magnetic field
generated in the upper dynamo active layer has to propagate
through a very thin layer only. So, it is reasonable to suggest that the spots produced in the latter case will be larger than those produced  in the former case.  Of course, this
suggestion needs to be confirmed by a dynamo model that would 
include a more or less realistic description of the spot formation.
There are various ideas concerning the particular mechanisms of formation of stellar spots  \citep[e.g.][]{P75, Betal2013, Jetal14, Getal16} so that
such confirmation requires an extensive modeling, which is, obviously, beyond the scope of this paper.
Another helpful point is that the G dwarfs where very strong flares occur are fast rotators, and one can expect that dynamo drivers here are substantially stronger than in the Sun. It may also help to produce more magnetic energy than on the Sun. Further modification of the idea is to suppose that the distribution of the dynamo drivers on superflaring stars differs substantially from that on the Sun (e.g., the activity wave propagates mainly towards the surface rather than towards the equator), which produces even more magnetic energy \citep{KO16, KK18, Ketal18}.

Note that the occurrence rate of weak flares (class C and even weaker
subflares)  is not related to the
spottedness. These weak flares occur in the Sun almost every day,
and their occurrence rate is almost independent on $A$ (Fig.~1).

We note that the points with error bars in Fig.~4 do not exactly approximate the data for particular stellar flares. It looks possible to improve this fitting introducing additional parameters in the model however our intention is to show that the general shape of flares distribution can be fitted by quite simple model and that various flares are related to compatible physical processes. We appreciate however that a further complication of the model may be interesting as well being associated with the fact that morphology of stellar spots may be different from solar ones.  In particular, we suppose here that the spots are more or less homogeneous. Supposing, that very powerful solar flares are associated with sunspot groups which contains many smaller sunspots what looks plausible according to available observations \cite{M90, Tetal19} one could obtain lower estimate for the point with $A=-3.0$ in Fig.~4.

To summarize our results, we can say that they are rather expected. Indeed, it looks plausible that larger spottendess gives more powerful stellar flares. Again, the very fact that the stellar activity cycles can be observed by contemporary observational methods means that there are stars with the spottendess substantially larger than observed in the Sun.  We note, however, that the expected results must be supported by detailed argumentation, which is presented in this paper.

Thus, the problem of a sharp difference in energy between the solar and stellar flares does not require revision of the flare models. This difference is due to the fact that the efficiency of spot formation decreases with the age of the star and the increase of its rotation period. This issue was studied in detail from a theoretical point of view in \citep{Pip21} (see also numerous references therein). Observational data also confirm this dependence for a large number of stars \citep{Tetal21}.
Here we concentrate attention mainly on G stars however M dwarfs data are obviously interested in this respect, see \cite{Netal17} and subsequent papers.

\begin{acknowledgments}
DDS thanks support by BASIS fund number 21-1-1-4-1. 
Authors thank the financial support of the Ministry of Science and Higher Education of the Russian Federation, program 075-15-2020-780 (VO, MK) and 075-15-2022-284 (DS). 
\end{acknowledgments}

\bibliography{sample631}{}
\bibliographystyle{aasjournal}

\end{document}